\RequirePackage{ifpdf}
\ifpdf 
\documentclass[pdftex]{sigma}
\else
\documentclass{sigma}
\fi

\numberwithin{equation}{section}

\begin{document}

\allowdisplaybreaks

\renewcommand{\thefootnote}{$\star$}

\renewcommand{\PaperNumber}{097}

\FirstPageHeading

\ShortArticleName{Symmetries of the Continuous and Discrete Krichever--Novikov Equation}

\ArticleName{Symmetries of the Continuous\\ and Discrete Krichever--Novikov Equation\footnote{This
paper is a contribution to the Proceedings of the Conference ``Symmetries and Integrability of Dif\/ference Equations (SIDE-9)'' (June 14--18, 2010, Varna, Bulgaria). The full collection is available at \href{http://www.emis.de/journals/SIGMA/SIDE-9.html}{http://www.emis.de/journals/SIGMA/SIDE-9.html}}}

\Author{Decio LEVI~$^\dag$, Pavel WINTERNITZ~$^\ddag$ and Ravil I.~YAMILOV~$^\S$}

\AuthorNameForHeading{D.~Levi, P.~Winternitz and R.I.~Yamilov}

\Address{$^\dag$~Dipartimento di Ingegneria Elettronica,   Universit\`a degli Studi Roma Tre and Sezione INFN,\\
\hphantom{$^\dag$}~Roma Tre,   Via della Vasca Navale 84, 00146 Roma, Italy}
\EmailD{\href{mailto:levi@roma3.infn.it}{levi@roma3.infn.it}}
\URLaddressD{\url{http://optow.ele.uniroma3.it/levi.html}}

\Address{$^\ddag$~Centre de recherches math\'{e}matiques and  D\'epartement de math\'ematiques et de statistique, \\
\hphantom{$^\ddag$}~Universit\'{e} de Montr\'{e}al, C.P.~6128, succ.\ Centre-ville, H3C 3J7, Montr\'{e}al (Qu\'{e}bec), Canada}
\EmailD{\href{mailto:wintern@crm.umontreal.ca}{wintern@crm.umontreal.ca}}
\URLaddressD{\url{http://www.crm.umontreal.ca/~wintern/}}

\Address{$^\S$~Ufa Institute of Mathematics, Russian Academy
of Sciences, \\
\hphantom{$^\S$}~112 Chernyshevsky Street, Ufa 450008, Russian Federation}
\EmailD{\href{mailto:RvlYamilov@matem.anrb.ru}{RvlYamilov@matem.anrb.ru}}
\URLaddressD{\url{http://matem.anrb.ru/en/yamilovri}}

\ArticleDates{Received June 16, 2011, in f\/inal form October 15, 2011;  Published online October 23, 2011}

\Abstract{A symmetry classif\/ication is performed for a class of dif\/ferential-dif\/ference equations depending on $9$ parameters. A $6$-parameter subclass
of these equations is an integrable discretization of the Krichever--Novikov equation. The dimension~$n$ of the Lie point symmetry algebra
satisf\/ies $1 \le n \le 5$. The highest dimensions, namely $n=5$ and $n=4$
occur only in the integrable cases.}


\Classification{35B06; 35K25; 
 37K10; 39A14}

\section{Introduction}\label{s1}

The Krichever--Novikov (KN) equation \cite{kn} is given by
\begin{gather} \label{KN}
\dot u = \frac{1}{4} u_{xxx} -\frac{3}{8} \frac{(u_{xx})^2}{u_x}+\frac{3}{2} \frac{P(u)}{u_x},\qquad \dot u \equiv u_t,
\end{gather}
where $P(u)$ is an arbitrary fourth degree polynomial of its argument with constant coef\/f\/icients. This is a nonlinear partial dif\/ferential equation with 5 arbitrary constant parameters.
Equation~(\ref{KN}) f\/irst appeared in the  study of the f\/inite gap solutions of the Kadomtsev--Petviashvili equation~\cite{1,kn,3}.  For a special choice of $P(u)$   (\ref{KN}) reduces to
 the Korteweg--de Vries equation but for a generic polynomial  no dif\/ferential substitution exists reducing equation~(\ref{KN}) to a KdV-type equation~\cite{4}. In \cite{kn, 5,9}, a zero-curvature
representation was obtained for~(\ref{KN}) involving  $sl(2)$ matrices. The Hamiltonian structure of
 (\ref{KN}) was analyzed and possible applications were reviewed in~\cite{6, 7}. B\"acklund transformations have been constructed together with the nonlinear superposition formulae in~\cite{a98}.  The Lax representation was used in~\cite{6} to
prove that~(\ref{KN}) has conservation laws.
In~\cite{bg10} the authors considered a generalization of~(\ref{KN}) in which the polyno\-mial~$P(u)$ is an arbitrary function of $u$ and studied its symmetry classif\/ication.

In 1983 Yamilov \cite{r5} introduced an integrable discretization of the Krichever--Novikov equation  (the YdKN equation):
\begin{gather}\label{e1}
\dot u_n \equiv u_{n,t} = f_n = \frac{S_n}{u_{n+1} - u_{n-1}} ,
\end{gather}
where the polynomial $S_n$ is given by
\begin{gather}
S_n  = P_n u_{n+1} u_{n-1} + Q_n (u_{n+1} + u_{n-1}) + R_n , \nonumber\\
P_n  =  \alpha u_n^2 + 2 \beta u_n + \gamma , \qquad
Q_n  =   \beta u_n^2 + \lambda u_n + \delta , \qquad
R_n  =  \gamma u_n^2 + 2 \delta u_n + \omega .\label{e2}
\end{gather}
This is a dif\/ferential-dif\/ference equation with 6 arbitrary constant parameters.
By carrying out the continuous limit,  we get the Krichever--Novikov equation  (\ref{KN})~\cite{r5} (see Section~\ref{section2} below).

The YdKN equation has been obtained as a result of a classif\/ication of dif\/ferential-dif\/ference equations  of the form $ \dot u_n = f(u_{n-1},u_n,u_{n+1})$ with no explicit $n$ and $t$ dependence \cite{r5,y06} that allow at least two conservation laws (or one conservation law and one generalized sym\-metry) of a~high enough order. In the general case, when all parameters are dif\/ferent from zero,  (\ref{e1}),~(\ref{e2})~is the only example in the complete list of Volterra type equations which cannot be transformed  by Miura transformations into the Volterra or Toda lattice equation~\cite{y06}. Recently it has been observed that most of the known integrable discrete equations on square lattices are closely related to the YdKN equation in the sense that they generate B\"acklund transformations of the YdKN equation \cite{lpsy,x09,ly11}. An $L{-}A$ pair for the YdKN equation has been constructed in~\cite{x09}.

A generalization of the YdKN equation (GYdKN)  introduced by D.~Levi and R.~Yamilov in~\cite{r51} has the same form (\ref{e1}), (\ref{e2}), but with $n$-dependent coef\/f\/icients:
\begin{gather}
P_n  =  \alpha u_n^2 + 2 \beta_n u_n + \gamma_n , \qquad
Q_n  =  \beta_{n+1} u_n^2 + \lambda u_n + \delta_{n+1} , \nonumber\\
R_n  =  \gamma_{n+1} u_n^2 + 2 \delta_n u_n + \omega .\label{e2*}
\end{gather}
Here $\beta_n$, $\gamma_n$, $\delta_n$ are two-periodic, i.e. can be written in the form
\begin{gather}\label{e2**}
\beta_n = \beta + \hat \beta (-1)^n , \qquad \gamma_n = \gamma + \hat \gamma (-1)^n , \qquad \delta_n = \delta + \hat \delta (-1)^n .
\end{gather}
Thus the GYdKN equation depends on 9 arbitrary constant parameters.
It has been shown in~\cite{r51} that the  GYdKN equation satisf\/ies the lowest integrability conditions in the generalized symmetry classif\/ication of Volterra type equations. Both YdKN and GYdKN equations are integrable in the sense that they possess master symmetries~\cite{asy} and therefore they have inf\/inite hierarchies of generalized symmetries and conservation laws. The GYdKN equation is also closely related to non-autonomous discrete equations on square lattices~\cite{xp09}.  It is worth mentioning here that this generalization does not allow a continuous limit to the Krichever--Novikov equation or any of its generalizations.

Extensions of the YdKN, which   in the continuous limit reduce to the KN equation or  its generalizations can be obtained by choosing $P_n$, $Q_n$ and $R_n$ as arbitrary $t$-independent functions of~$u_n$.
An interesting    extension of the YdKN equation is given by the equation
\begin{gather} \label{4.1}
u_{n,t}  \equiv \dot u_n  =  \frac{P(u_n) u_{n+1} u_{n-1} + Q(u_n) (u_{n+1} + u_{n-1}) + R(u_n)}{u_{n+1} - u_{n-1}}, \\
P(u_n)  =  \alpha u_n^2 + 2 \beta u_n + \gamma, \qquad\!
Q(u_n)  =  \hat \beta u_n^2 + \lambda u_n + \delta, \qquad\!
R(u_n)  =  \hat \gamma u_n^2 + 2 \hat \delta u_n  + \omega,\!\!\!\label{4.2}
\end{gather}
where $\alpha, \dots, \omega$ are 9 real constants, at least one of them nonzero. We will call~(\ref{4.1})  the EYdKN (extended YdKN). Like  the GYdKN the EYdKN equation depends on~9 constant coef\/f\/icients. By choosing $\hat \beta=\beta$, $\hat \gamma=\gamma$, $\hat \delta=\delta$ it reduces to the YdKN equation.

In the following  we are going to carry out the point symmetry classif\/ication for all particular cases of the EYdKN   equation. These are dif\/ferential-dif\/ference equations and for them we will use the theory of symmetries of dif\/ference equations as presented in~\cite{r2,r4,r42,r31,lwy2010}.  Due to its complication we present here just one example of an equation belonging to the GYdKN class which possesses a nontrivial point symmetry algebra.

In Section~\ref{section2} we f\/irst take the continous limit of a generalized YdKN equation and then calculate the Lie point symmetries of the obtained (continuous) generalized Krichever--Novikov equation (\ref{KN}) in which $f(u) \equiv P(u)$ is an arbitrary function. Sections~\ref{section3} and~\ref{section4} are devoted to a~symmetry classif\/ication of the EYdKN equation for which $P(u_n)$, $Q(u_n)$ and $R(u_n)$ are restricted to being second order polynomials. This includes the integrable YdKN equation as a~subcase.  Some conclusions and future outlook are presented in Section 5.

\section{Continuous limit of a generalized YdKN equation\\ and its Lie point symmetries}\label{section2}

\subsection{The continuous limit}\label{section2.1}

Let us look for the continuous limit of a generalization of the YdKN equation~(\ref{e1}), (\ref{e2}). Here, for the sake of simplicity of notation we take
$P(u_n)$, $Q(u_n)$ and $R(u_n)$
 as arbitrary functions of their argument. We carry out the continuous limit generalizing the procedure used in \cite{r5}.

First of all we redef\/ine the functions $P(u_n)$, $Q(u_n)$ and $R(u_n)$
\begin{gather*} 
P(u_n)=\tilde P(u_n)+ k ,\qquad Q(u_n)=\tilde Q(u_n) - k u_n, \qquad R(u_n) = \tilde R(u_n) + k u_n^2,
\end{gather*}
where $k$ is  an arbitrary constant.
We introduce a small parameter $h$, the lattice spacing, by putting
\begin{gather*}
 \tilde P(u_n) = 2hF(v(x,t)), \qquad  \tilde Q(u_n) = 2h G(v(x,t)), \qquad \tilde R(u_n) = 2h H(v(x,t)), \nonumber\\  
 u_n(t)=v(x,t), \qquad x=nh+6\frac{t}{h^2}, \qquad k=-\frac{12}{h^3}.
\end{gather*}
Taking the limit,  $h \rightarrow 0$ and $n \rightarrow \infty$  with $nh$ f\/inite,  we get
\begin{gather*} 
v_t =  v_{xxx} - \frac{3}{2} \frac{v_{xx}^2}{v_x} +\frac{v^2 F(v)+ 2 v G(v)+ H(v)}{v_x} + \mathcal O\big(h^2\big).
\end{gather*}
Putting $v^2 F(v) + 2 v G(v) + H(v) = f(v)$ and replacing $v(x,t)$ by $u(x,t)$ we obtain the ``generalized Krichever--Novikov equation''
\begin{gather} \label{c1}
u_t = u_{xxx} - \frac{3}{2} \frac{(u_{xx})^2}{u_x} + \frac{f(u)}{u_x}.
\end{gather}
Rescaling and restricting the arbitrary function $f(u)$ to a fourth order polynomial we obtain the Krichever--Novikov equation~(\ref{KN}).

\subsection[Lie point symmetries of the continuous generalized Krichever-Novikov equation]{Lie point symmetries of the continuous\\ generalized Krichever--Novikov equation}\label{section2.2}

For comparison with the extended YdKN equation (\ref{4.2}) we present a symmetry classif\/ication of~(\ref{c1}), thus completing the partial classif\/ication performed in~\cite{bg10}.

Equation~(\ref{c1}) is form-invariant under ``allowed transformations'' that only change the form of $f(u)$. These include M\"obius transformations of the dependent variable $u$ and a simultaneous rescaling of $x$ and $t$
\begin{gather} \label{c2}
u=\frac{\alpha \tilde u + \beta}{\gamma \tilde u + \delta}, \qquad \alpha \delta - \beta \delta \ne 0,  \qquad t=k^3 \tilde t, \qquad x=k \tilde x.
\end{gather}
The function $f(u)$ transforms into
\begin{gather} \label{c3}
\tilde f(\tilde u)= \frac{k^4}{(\alpha \delta - \beta \gamma)^2} f\left(\frac{\alpha u + \beta}{\gamma u + \delta}\right)(\gamma u + \delta)^4.
\end{gather}
We shall classify  (\ref{c1}) into symmetry classes under the action of the group of allowed transformations (\ref{c2}). The ``group of allowed transformations'' is sometimes also called the ``equivalence group'' of the equation.

We write a general element of the symmetry algebra of  (\ref{c1}) in the form
\begin{gather} \label{c3a}
X = \tau(x,t,u) \partial_t + \xi(x,t,u) \partial_x + \phi(x,t,u) \partial_u.
\end{gather}
Requiring that the third prolongation ${\rm pr}^{(3)} X$ of (\ref{c3a}) should annihilate  (\ref{c1}) on the solution set, we obtain $9$ determining equations for the coef\/f\/icients $\tau$, $\xi$ and $\phi$.  The f\/irst $8 $ of them are elementary and imply
\begin{gather*} 
\tau = \tau_1 t + \tau_0, \qquad \xi = \frac{1}{3}\tau_1 x + \xi_0, \qquad \phi=\phi_2 u^2 + \phi_1 u + \phi_0,
\end{gather*}
where $\tau_0$, $\tau_1$, $\xi_0$, $\phi_2$, $\phi_1$ and $\phi_0$ are constants. The remaining determining equation implies that the function $f(u)$ f\/iguring in  (\ref{c1}) must satisfy the following f\/irst order ODE:
\begin{gather} \label{c5}
\big(\phi_2 u^2 + \phi_1 u + \phi_0\big) \frac{df}{du} + \left[-4 \phi_2 u + \frac{4}{3} \tau_1 - 2 \phi_1 \right] f =0.
\end{gather}
A symmetry analysis of  (\ref{c1}) thus boils down to analyzing all possible solutions of (\ref{c5}).

First of all  (\ref{c5}) does not contain $\tau_0$ and $\xi_0$. This is just a ref\/lection of the obvious fact that  (\ref{c1}) does not depend explicitly on $t$ and $x$ and is hence invariant under time and space  translations for any function $f(u)$. They are generated by
\begin{gather} \label{c6}
P_0 = \partial_t, \qquad P_1 = \partial_x,
\end{gather}
respectively.

Let us now assume that at least one of the coef\/f\/icients  $\phi_0$, $\phi_1$, $\phi_2$ or $\tau_1$ is nonzero.
In Table~\ref{table1} we present representatives of all classes of functions~$f(u)$ for which the symmetry algebra~$L$ of~(\ref{c1}) is larger than (\ref{c6}). We have $2 < \dim L \le 6$ in all cases. The classif\/ication is under the allowed transformations~(\ref{c2}),~(\ref{c3}). The following cases occur, depending on the properties of the polynomial
\begin{gather*} 
\phi(u)=\phi_2 u^2 + \phi_1 u + \phi_0.
\end{gather*}
\begin{enumerate}\itemsep=0pt

\item $\phi_2 \ne 0$, $\phi(u)=0$ has complex roots $u_{1,2}=r \pm i s$, $s > 0$. After an allowed transformation the solution is
    \begin{gather*} 
    f(u)=f_0\big(1+u^2\big)^2 e^{p \arctan u}, \qquad f_0 = \pm 1,
    \end{gather*}
    with $p \in \mathbb R$;  $p=0$ is a special case.

\item $\phi_2 \ne 0$, $\phi(u)=0$ has two real roots  $u_1 < u_2$,
\begin{gather} \label{c9}
f(u)=f_0(u+1)^p (u-1)^{4-p}, \qquad f_0 = \pm 1.
\end{gather}
Since  $p$ and $4-p$ are equivalent, we can restrict to the case $2 \le p < \infty$. The case $p=2$ is again special. For $p=2, 3, 4$ (\ref{c9}) is a fourth order polynomial.

\item $\phi_2 \ne 0$, $\phi(u) =0$ has a double root and we shift it to $u_1=u_2=0$. We obtain $f(u)=f_0 u^4 e^{-\frac{p}{u}}$. An allowed transformation takes this into
    \begin{gather} \label{c10}
    f(u) = f_0 e^u, \qquad f_0 = \pm 1.
    \end{gather}

\item $\phi_2 =0$, $\phi_1 \ne 0$ We obtain
\begin{gather*} 
f(u) = f_0 u^p, \qquad f_0 = \pm 1.
\end{gather*}
Under an allowed transformation we have $p \rightarrow 4-p$ so we can restrict $p$ to $2 \le p < \infty$. The case $p=4$ is equivalent to $f(u) = f_0$.

\item $\phi_2=0$, $\phi_1=0$, $\phi_0 \ne 0$. We reobtain (\ref{c10}) or $f(u)=f_0$.
\end{enumerate}
In Table~\ref{table1} we give the functions $f(u)$ in column~2 and the basis elements of the Lie algebra in column~4. Throughout we have $f_0 = \pm 1$ and we use the notation (\ref{c6}) and
\begin{gather*} 
D = t \partial_t + \frac{1}{3} x \partial_x,\qquad U_0 = \partial_u, \qquad U_1=u \partial_u, \qquad U_2 = u^2 \partial_u.
\end{gather*}
The Lie algebras in cases $3$, $4$, $5$ and $6$ of Table $1$ are all solvable with $\{ P_0, P_1 \}$ as their nilradical. The values $p=2$ in case $3$ and $6$ and $p=0$ in case 5 are special as the Lie algebra for these values contracts to an Abelian one.
\begin{table}[t]\centering
\caption{Symmetry classif\/ication of the continuous generalized Krichever--Novikov equation; $f_0 = \pm 1$, see Section~\ref{section2} for notation.}\label{table1}
\vspace{1mm}

\begin{tabular}{|c|c|c|c|}
\hline
$N_0$ & $f(u)$& dim $L$ & Basis elements of symmetry algebra $L$ \tsep{1pt}\bsep{1pt}\\ \hline \hline
1 & 0 & 6 & $P_0$, $P_1$, $D$, $U_0$, $U_1$, $U_2$\tsep{1pt}\bsep{1pt}\\ \hline
2 & $f_0$ & 4 &$P_0$, $P_1$, $D+\frac{2}{3}U_1$, $U_0$ \tsep{1pt}\bsep{1pt}\\ \hline
 3 & $f_0 u^p$ & 3 & $P_0$, $P_1$, $(p-2)D - \frac{4}{3}  U_1$, $2 \le p<\infty$ \tsep{1pt}\bsep{1pt}\\ \hline
4 & $f_0 e^u$ & 3 & $P_0$, $P_1$, $D - \frac{4}{3} U_0$ \tsep{1pt}\bsep{1pt}\\ \hline
 5 & $f_0 (u^2+1)^2 e^{p \arctan u}$ & 3 & $P_0$, $P_1$, $pD - \frac{4}{3}(U_2 + U_0)$ \tsep{1pt}\bsep{1pt}\\ \hline
 6 & $f_0 (u+1)^p (u-1)^{4-p}$ & 3 & $P_0$, $P_1$, $(p-2)D - \frac{2}{3}(U_2 - U_0)$ $2 \le p<\infty$\tsep{1pt}\bsep{1pt} \\ \hline
\end{tabular}
\end{table}

\section{Symmetry structure of the extended YdKN equation}\label{section3}

\subsection{Allowed transformations}\label{section3.1}

First of all we notice that  (\ref{4.1}), (\ref{4.2}) is form-invariant under the M\"obius transformation
\begin{gather} \label{b2.1}
u_n \rightarrow \tilde u_n = \frac{\eta_1 u_n + \eta_2}{\eta_3 u_n + \eta_4}, \qquad \Delta = \eta_1 \eta_4 - \eta_2 \eta_3 = \pm 1,
\end{gather}
where $\eta_i$, $i=1, \dots, 4$ are arbitrary real constants. All such $SL(2,\mathbb R)$ transformations can be induced by combinations of translations $\tilde u_n = u_n +\kappa$, dilations $\tilde u_n = \kappa u_n $ and the inversion $\tilde u_n = 1/u_n$. Explicitly the coef\/f\/icients $\alpha, \dots, \omega$ of~(\ref{4.2}) transform under a translation into
\begin{gather}
 \alpha^*=\alpha, \qquad \beta^* = \beta +\alpha \kappa, \qquad \hat \beta^* = \hat \beta + \alpha \kappa, \nonumber
\qquad
 \gamma^*=\gamma + 2 \beta \kappa + \alpha \kappa^2,\nonumber\\  \hat \gamma^*=\hat \gamma + 2 \hat \beta \kappa + \alpha \kappa^2, \qquad
 \lambda^*=\lambda + 2 (\beta + \hat \beta)\kappa + 2 \alpha \kappa^2,  \nonumber\\ \nonumber
 \hat \delta^*=\hat \delta +(\hat \gamma + \lambda) \kappa+(\beta + 2 \hat \beta) \kappa^2 + \alpha \kappa^3, \qquad
  \delta^*= \delta +( \gamma + \lambda) \kappa+(2 \beta +  \hat \beta) \kappa^2 + \alpha \kappa^3, \\
 \omega^*=\omega +2(\delta + \hat \delta) \kappa + (\gamma + \hat \gamma + 2 \lambda) \kappa^2 +2 (\beta +  \hat \beta) \kappa^3 + \alpha \kappa^4, \label{b2.2}
\end{gather}
under a dilation into
\begin{gather*} \nonumber
 \alpha^*=\alpha \kappa^2, \qquad \beta^*=\beta \kappa, \qquad  \hat \beta^*=\hat \beta \kappa, \qquad
 \gamma^*=\gamma, \qquad \hat \gamma^*=\hat \gamma, \qquad   \lambda^*= \lambda, \\
 \delta^*=\delta/\kappa, \qquad \hat \delta^*=\hat \delta/\kappa, \qquad   \omega^*= \omega/\kappa^2,
\end{gather*}
and under the inversion into
\begin{gather*} \nonumber
 \alpha^*=\omega, \qquad \beta^*=\hat \delta, \qquad  \hat \beta^*= \delta, \qquad
 \gamma^*=\hat \gamma, \qquad \hat \gamma^*= \gamma, \qquad   \lambda^*= \lambda, \\
 \delta^*=\hat \beta, \qquad \hat \delta^*= \beta, \qquad   \omega^*= \alpha.
\end{gather*}
Equation~(\ref{4.1}) is also form-invariant under dilation of time (and  invariant under time translation).

\subsection{Theorems simplifying the symmetry classif\/ication}\label{section3.2}

First of all, we shall show that we can restrict the study of the EYdKN equation to the case $P(u_n) \ne 0$ in~(\ref{4.2}) and that this can be split into precisely 3 subcases.

\begin{theorem}\label{2t1}\sloppy
 Using the M\"obius transformation \eqref{b2.1}  we can reduce the EYdKN equa\-tion~\eqref{4.1},~\eqref{4.2} for $(P_n, Q_n, R_n) \ne (0,0,0)$ to one of the $3$ following cases:
\begin{enumerate}\itemsep=0pt
 \item[$1.$] $\alpha=1$, $\beta=0$;
\item[$2.$] $\alpha=0$, $\beta=1$, $\gamma=0$;
\begin{gather} \label{b2.5}
\beta + \hat \beta = \delta + \hat \delta = \gamma + \hat \gamma + 2 \lambda = \omega =0;
\end{gather}
\item[$3.$] $\alpha=0$, $\beta=0$, $\gamma=1$
\begin{gather} \label{b2.6}
 \hat \beta = \delta =\hat \delta = \gamma + \lambda=\hat \gamma +  \lambda = \omega =0.
\end{gather}
\end{enumerate}
In all cases we have $P(u_n) \ne 0$.
\end{theorem}
\begin{proof}
 If $\alpha \ne 0$ we can scale it to $\alpha=1$ and then transform $\beta$ into $\beta=0$ by a translation of~$u_n$.

 Now assume $\alpha=0$, $\beta \ne 0$.
  Up to M\"obius transformations we must also assume $\alpha^*=0$ in (\ref{b2.2}). This imposes the conditions (\ref{b2.5}) on the other coef\/f\/icients in the EYdKN equation, otherwise we can always chose $\kappa$ to obtain $\alpha^* \ne 0$.
For $\beta \ne 0$ we can again dilate  to obtain $\beta=1$ and translate~$u_n$ to obtain~$\gamma=0$.

The third case corresponds to $\alpha=\beta=0$, $\gamma \ne 0$ and we dilate to obtain $\gamma=1$. Conditions~(\ref{b2.6}) follow from the requirement $\alpha^*=0$, $\beta^*=0$ for all values of $\eta_i$, $i=1, \dots, 4$ in the M\"obius transformation.

Finally, if we impose $\alpha=\beta=\gamma=0$ and also $\alpha^*=\beta^*=\gamma^*=0$ for all values $\eta_i$, $i=1, \dots, 4$ we obtain not only $P(u_n)=0$ but also $Q(u_n)=R(u_n)=0$, i.e.~(\ref{4.1}) is trivial.
\end{proof}

\noindent
{\bf Comment.} A further scaling of one more parameter  can be achieved using a dilation of time~$t$. This can provide simplif\/ications which will be discussed below in Section~\ref{section4} in each specif\/ic case.

\begin{theorem}\label{2t2}\sloppy
The Lie algebra of local Lie point symmetries of the EYdKN equation  with $(P_n, Q_n, R_n) \ne (0,0,0)$ consists of vector fields of the general form
\begin{gather} \label{b2.7}
X = \tau(t) \partial_t + \phi_n(t,u_n) \partial_{u_n}
\end{gather}
with
\begin{gather} \label{b2.8}
 \tau = \tau_0 + \tau_1 t, \qquad \phi_n = a_n + b_n u_n + c_n u_n^2, \\ \label{b2.9}
 a_n = a + \hat a (-1)^n, \qquad b_n = b + \hat b (-1)^n, \qquad c_n = c + \hat c (-1)^n,
\end{gather}
where $\tau_0$, $\tau_1$, $a$, $\hat a$, $b$, $\hat b$, $c$ and $\hat c$ are constants.
\end{theorem}

\begin{proof}
 In a previous article \cite{lwy2010} we have shown that for a large class of dif\/ferential-dif\/ference equations, including the EYdKN equation, the vector f\/ield corresponding to Lie point symmetries must have the form (\ref{b2.7}), in particular $\tau(t)$ does not depend on~$n$ or~$u_n$. The f\/irst prolongation of~$X$ to be applied to (\ref{4.1}) is
\begin{gather*} 
{\rm pr}\,X = \tau(t) \partial_t  + \sum_{j=n-1}^{n+1} \phi_j(t,u_j) \partial_{u_j} + \phi_n^{(1)} \partial_{\dot u_n}, \qquad
\phi_n^{(1)} =  D_t \phi_n(t,u_n) - [D_t \tau(t) ] \dot u_n,
\end{gather*}
where $D_t$ is the total derivative operator. Applying $\mbox{pr}\,X$ to the equation and requiring the result to be zero on the solution set, we obtain the determining equation
\begin{gather} \nonumber
  \phi_{n,t} (u_{n+1}-u_{n-1})^2  +(\phi_{n,u_n}-\dot \tau) [ P u_{n+1} u_{n-1} + Q (u_{n+1} + u_{n-1}) + R](u_{n+1}-u_{n-1})  \\ \nonumber
\qquad{} - \phi_n [P_{,u_n} u_{n+1} u_{n-1} + Q_{,u_n} (u_{n+1} + u_{n-1}) + R_{,u_n}](u_{n+1}-u_{n-1})   \\
\qquad {} = \phi_{n-1} \big[ P u_{n+1}^2 + 2 Q u_{n+1} +R\big] - \phi_{n+1}\big[P u_{n-1}^2 + 2 Q u_{n-1} +R\big].\label{b2.11}
\end{gather}
From Theorem~\ref{2t1} we know that we only need to consider the case $P(u_n)\ne 0$. Applying the fourth derivative $\partial^2_{u_{n+1}} \partial^2_{u_{n-1}} $ to (\ref{b2.11}) and dividing by $P_n$ we obtain
\begin{gather*} 
\phi_{n+1,u_{n+1}u_{n+1}} - \phi_{n-1,u_{n-1}u_{n-1}}=0.
\end{gather*}
This implies
\begin{gather} \label{b2.13}
\phi_n = a_n + b_n u_n + c_n u_n^2, \qquad c_n = c + \hat c (-1)^n.
\end{gather}
Putting (\ref{b2.13}) back into (\ref{b2.11}) and comparing independent terms we obtain
\begin{gather*} 
\ddot \tau = 0, \qquad a_{n+1}=a_{n-1}, \qquad b_{n+1}=b_{n-1},
\end{gather*}
where all coef\/f\/icients are time independent. This completes the proof of Theorem~\ref{2t2}.
\end{proof}

\subsection{The determining equations}\label{section3.3}

Let us now return to (\ref{b2.11}) and substitute into it the expression (\ref{b2.8}) and (\ref{b2.9}) for $\tau$ and $\phi_n$, as well as (\ref{4.2}) for $P$, $Q$ and $R$. The expressions multiplying $(u_{n+1})^k (u_{n-1})^{\ell}$ for dif\/ferent values of $k$ and $\ell$ must vanish separately. This will provide us with two sets of linear algebraic homogeneous equations, one for the {\it vector} $\vec v_1=(a,b,c,\tau_1)$, the other for $\vec v_2=(\hat a,\hat b,\hat c)$. The coef\/f\/icient $\tau_0$ does not f\/igure anywhere, so
\begin{gather*} 
P_0 = \frac{\partial}{\partial t}
\end{gather*}
is always an element of the algebra.

We write these two {\it matrix determining equations} as
\begin{gather*}
\hat M_1 \vec v_1 = \vec 0, \qquad \hat M_2 \vec v_2 = \vec 0
\end{gather*}
and denote
\begin{gather*} 
r_1= \mbox{rank}\, \hat M_1, \qquad r_2= \mbox{rank}\, \hat M_2.
\end{gather*}
The dimension of the symmetry algebra of the EYdKN equation will be
\begin{gather*}
\dim  L = 8 - (r_1 + r_2),
\end{gather*}
and in view of Theorem \ref{2t1} we need only to consider the case $P \ne 0$. The two matrices involved are
\begin{gather} \label{b2.19}
\hat M_{1}  =  \left(
\begin{array}{cccc}
\alpha & \beta & -(\lambda+\gamma) & \beta\\
\alpha &\hat  \beta & -(\lambda+\hat \gamma) & \hat \beta \\
0 & 2 \alpha &-2 (\beta + \hat \beta) & \alpha \\
2 \beta & 0 & - 2 \delta & \gamma \\
-2(\beta+\hat \beta) & 0 & 2(\delta + \hat \delta) & - \lambda \\
2 \hat \beta & 0  & - 2 \hat \delta & \hat \gamma \\
-(\lambda+\gamma) & \delta & \omega &-\delta \\
-(\lambda+\hat \gamma) & \hat \delta & \omega &-\hat \delta \\
2(\delta+\hat \delta) & -2\omega & 0 & \omega
\end{array}\right), \\ \label{b2.20}
\hat M_{2}  =  \left(
\begin{array}{ccc}
\alpha & -\beta & -(\gamma-\lambda) \\
\alpha & - \hat \beta & -(\hat \gamma-\lambda)  \\
\beta & - \gamma & \delta \\
\hat \beta & - \hat \gamma & \hat \delta \\
\beta-\hat \beta & 0 & \delta - \hat \delta  \\
\gamma-\lambda & \delta & -\omega  \\
\hat \gamma-\lambda & \hat \delta & -\omega  \\
\delta-\hat \delta &  0 & 0 \\
0 & 0 & \beta - \hat \beta
\end{array}\right).
\end{gather}

\section{Symmetry classif\/ication for the EYdKN equation}\label{section4}

\subsection{General comments}\label{section4.1}

Let us introduce a notation for the matrix of coef\/f\/icients of the EYdKN equation (\ref{4.1}), (\ref{4.2}):
\begin{gather*} 
K  =  \left(
\begin{array}{ccc}
\alpha & \beta & \gamma \\
\hat \beta & \lambda & \delta  \\
\hat \gamma  &  \hat \delta & \omega
\end{array}\right).
\end{gather*}
We will analyze the possible ranks of the two matrices $\hat M_1$ and $\hat M_2$ of (\ref{b2.19}) and (\ref{b2.20}), as functions of the coef\/f\/icients in~$K$. We shall f\/irst determine all cases when the ranks $r_1$ and $r_2$ satisfy $r_1+r_2 \le 5$ so that the symmetry algebra $L$ has dimension $3 \le \dim L \le 5$. Separately we list all cases when we have $\dim L=2$.

\subsection[Symmetry algebras with $\dim L \ge 3$ and $\alpha \ne 0$]{Symmetry algebras with $\boldsymbol{\dim L \ge 3}$ and $\boldsymbol{\alpha \ne 0}$}\label{section4.2}

We take $\alpha=1$, $\beta=0$. Inspecting the matrices $\hat M_1$ and $\hat M_2$ of (\ref{b2.19}) and (\ref{b2.20}) we see that their ranks satisfy
\begin{gather*} 
2 \le r_1 \le 4, \qquad 1 \le r_2 \le 3.
\end{gather*}
The cases relevant for this section are:

{\bf $\boldsymbol{\dim L=5}$.}

$r_1=2$, $r_2=1$, $\alpha=1$, $\beta=\hat \beta=\gamma=\hat \gamma=\delta=\hat \delta=\lambda=\omega=0$.
\begin{gather} \label{b4.3}
K  =  \left(
\begin{array}{ccc}
1 & 0 & 0 \\
0 & 0 & 0  \\
0   &  0 & 0
\end{array}\right),\qquad \dot u_n = \frac{u_n^2 u_{n+1} u_{n-1}}{u_{n+1}-u_{n-1}},\\
\nonumber
X_0  =  \partial_t, \qquad X_1=2 t \partial_t-u_n\partial_{u_n}, \qquad X_2= (-1)^n u_n\partial_{u_n}, \\ \nonumber
X_3  =  u_n^2 \partial_{u_n}, \qquad X_4 = (-1)^n u_n^2\partial_{u_n}.
\end{gather}
This is a 5-dimensional solvable Lie algebra with Abelian nilradical $\{ X_0, X_3, X_4 \}$. The nonnilpotent elements $\{X_1, X_2 \}$ commute  and have a diagonal action on the nilradical. We mention that the allowed transformation $u_n \rightarrow \frac{1}{u_n}$ takes (\ref{b4.3}) into the dif\/ferential-dif\/ference equation
\begin{gather*} 
\dot u_n = \frac{1}{u_{n+1}-u_{n-1}}
\end{gather*}
 sometimes called the discrete KdV equation \cite{nc95,lp2007,lpsy,x2011,rh2007}.

 {\bf  $\boldsymbol{\dim L=4}$.}

$r_1=3$, $r_2=1$, $\alpha=1$, $\beta=\hat \beta=\gamma=\hat \gamma=\delta=\hat \delta=0$, $\omega=\lambda^2$, $\lambda= \pm 1$.
\begin{gather} \label{b4.4}
K  =  \left(
\begin{array}{ccc}
1 & 0 & 0 \\
0 & \lambda & 0  \\
0   &  0 & 1
\end{array}\right),\qquad \dot u_n = \frac{u_n^2 u_{n+1} u_{n-1}+\lambda u_n ( u_{n+1}+u_{n-1}) + \lambda^2}{u_{n+1}-u_{n-1}},\\ \nonumber
X_0  =  \partial_t, \qquad X_1=(-1)^n u_n\partial_{u_n}, \qquad X_2= (-1)^n (u_n^2 - \lambda)\partial_{u_n},\qquad  X_3 = (u_n^2 + \lambda) \partial_{u_n}.
\end{gather}
The algebra is reductive and isomorphic to $gl(2, \mathbb R)$.

 {\bf  $\boldsymbol{\dim L=3}$.} We obtain two cases, namely:
\begin{enumerate}\itemsep=0pt
\item $r_1=3$, $r_2=2$, $\alpha=1$, $\beta=\hat \beta=\delta=\hat \delta=\lambda=\omega=0$, $\gamma=\hat \gamma= \pm 1$.
\begin{gather}
K  =  \left(
\begin{array}{ccc}
1 & 0 & \gamma \\
0 & 0 & 0 \\
\gamma   &  0 & 0
\end{array}\right), \qquad
\dot u_n = \frac{(u_n^2 +\gamma)(u_{n+1} u_{n-1}+\gamma)}{u_{n+1}-u_{n-1}},\nonumber\\
X_0  =  \partial_t,  \quad X_1= (-1)^n \big(u_n^2 + \gamma\big)\partial_{u_n},\quad  X_2 = \big(u_n^2 + \gamma\big) \partial_{u_n}.\label{b4.5}
\end{gather}
The algebra is Abelian.

\item $r_1=3$, $r_2=2$, $\alpha=1$, $\beta=\hat \beta=0$, $\gamma=\hat \gamma= -\mu^2$, $\delta=\hat \delta=2 \mu^3$, $\lambda=-2 \mu^2$ and $\omega=-3 \mu^4$.
\begin{gather}
K   =  \left(
\begin{array}{ccc}
1 & 0 & -\mu^2 \\
0 & -2 \mu^2 & 2 \mu^3 \\
- \mu^2  &  2 \mu^3 & -3 \mu^4
\end{array}\right), \nonumber\\  \dot u_n  =  \frac{(u_n-\mu)[(u_n +\mu)u_{n+1} u_{n-1} - 2\mu^2(u_{n+1} +u_{n-1})-\mu^2(u_n-3\mu)]}{u_{n+1}-u_{n-1}},\label{b4.5a}\\
X_0  =  \partial_t,  \qquad X_1= (-1)^n (u_n - \mu)^2\partial_{u_n},\qquad  X_2 =t \partial_t-\frac{1}{4 \mu} (u_n+3 \mu)(u_n - \mu ) \partial_{u_n}.\nonumber
\end{gather}
The algebra is solvable with an Abelian nilradical $\{ X_0, X_1 \}$.
\end{enumerate}

\subsection[Symmetry algebras with $\dim L \ge 3$ and $\alpha=0$, $\beta \ne 0$]{Symmetry algebras with $\boldsymbol{\dim L \ge 3}$ and $\boldsymbol{\alpha=0}$, $\boldsymbol{\beta \ne 0}$}\label{section4.3}

In this case we take $\beta=1$, $\gamma=0$ and in view of (\ref{b2.5}) we have
\begin{gather*}
 \alpha=0, \qquad \beta=1, \qquad \gamma=0, \qquad \hat \beta=-1, \qquad \hat \delta=-\delta, \qquad \hat \gamma = -2 \lambda, \qquad \omega=0,
 \end{gather*}
so the matrix $K$ is
\begin{gather*} 
K  =  \left(
\begin{array}{ccc}
0 & 1 & 0 \\
-1 & \lambda & \delta  \\
-2 \lambda  &  - \delta & 0
\end{array}\right).
\end{gather*}
In this case the rank of $\hat M_2$ is always $r_2=3$ so we have $\hat a= \hat b= \hat c =0$ in (\ref{b2.9}). The dimension of the symmetry algebra is $\dim L=5-r_1$.
The only case of interest here is $r_1=2$ and that requires $\lambda=\delta=0$. In this case we have

 {\bf  $\boldsymbol{\dim L=3}$.}

$r_1=2$, $r_2=3$. $\beta=1$, $\hat \beta=-1$, $\alpha=\gamma=\hat \gamma=\delta=\hat \delta=\lambda=\omega=0$.
\begin{gather} \label{b4.8}
 K  =  \left(
\begin{array}{ccc}
0 & 1 & 0 \\
-1 & 0 & 0  \\
0  &  0 & 0
\end{array}\right),\quad \dot u_n = \frac{2 u_n u_{n+1} u_{n-1}- u_n^2( u_{n+1} + u_{n-1})}{u_{n+1}-u_{n-1}},\\ \nonumber
X_0  =  \partial_t,  \qquad X_1= t \partial_t -u_n \partial_{u_n},\qquad  X_2 = u_n^2  \partial_{u_n}.
\end{gather}
The algebra is solvable with an Abelian nilradical $\{X_0,X_2\}$.

\subsection[Symmetry algebras with $\dim L \ge 3$ and $\alpha=\beta=0$, $\gamma \ne 0$]{Symmetry algebras with $\boldsymbol{\dim L \ge 3}$ and $\boldsymbol{\alpha=\beta=0}$, $\boldsymbol{\gamma \ne 0}$}\label{section4.4}

We normalize $\gamma$ to $\gamma=1$ by rescaling $t$ and have
\begin{gather*} 
K  =  \left(
\begin{array}{ccc}
0 & 0 & 1 \\
0 & -1 & 0  \\
1   &  0 & 0
\end{array}\right).
\end{gather*}
This leads to one further four-dimensional Lie algebra, namely

 {\bf  $\boldsymbol{\dim L=4}$.}

$r_1=1$, $r_2=3$.
\begin{gather} \label{b4.11}
  \dot u_n = \frac{ u_{n+1} u_{n-1}- u_n( u_{n+1} + u_{n-1})+ u_n^2}{u_{n+1}-u_{n-1}},\\ \nonumber
 X_0 = \partial_t,  \qquad X_1=  \partial_{u_n},\qquad X_2= u_n \partial_{u_n},\qquad  X_3 = u_n^2  \partial_{u_n}.
\end{gather}
The algebra is isomorphic to $gl(2, \mathbb R)$.

\subsection[Symmetry algebras of dimension $\dim L=2$]{Symmetry algebras of dimension $\boldsymbol{\dim L=2}$}\label{section4.5}

A symmetry algebra of dimension $\dim L=2$ will have one element $X$, in addition to $X_0= \partial_t$.

The element $X$ can have one of two forms: $X=(a + b u_n + c u_n^2) \partial_{u_n} + \tau_1 \partial_t$ and it occurs for $r_1=3$, $r_2=3$, or $X=(-1)^n(\hat a + \hat b u_n + \hat c u_n^2) \partial_{u_n}$ for $r_1=4$, $r_2=2$. We shall consider the two cases separately, following the same branches as for $\dim L\ge 3$.

\subsubsection*{I. $\boldsymbol{r_1=3}$, $\boldsymbol{r_2=3}$}

 {\bf Branch 1.} $\boldsymbol{\alpha=1}$, $\boldsymbol{\beta=0}$

 {\bf $\boldsymbol{{\bf I}_1}$, $\boldsymbol{\gamma \ne 0}$} (we can normalize it to $\gamma = \pm 1$).
The matrix $\hat M_1$ is equivalent to
\begin{gather*} 
\hat M_1  \sim  \left(
\begin{array}{cccc}
1 & 0 & -(\lambda+\gamma) & 0\\
0 & 1 & -\hat \beta & \frac{1}{2}  \\
0   &  0 & -\frac{2\delta}{\gamma} & 1
\end{array}\right)
\end{gather*}
with all further rows vanishing in order for the rank to be $r_1=3$. This implies the following conditions  on the parameters in the equation
\begin{gather}
\gamma - \hat \gamma + \hat \beta^2 + \frac{\delta \hat \beta}{\gamma} = 0, \qquad
  \delta + \hat \delta - \hat \beta (\lambda + \gamma) -  \frac{\lambda \delta}{\gamma} = 0, \nonumber\\
 \omega - (\lambda + \gamma)^2 + \hat \beta \delta - 3 \frac{ \delta^2}{\gamma} = 0, \qquad
 \omega - (\lambda+\gamma)(\lambda+\hat \gamma) + \hat \beta \delta - 3 \frac{\delta \hat \delta}{\gamma} = 0,\nonumber \\
 (\delta+\hat \delta)(\lambda+\gamma) - \hat \beta \omega + \frac{2 \delta \omega}{\gamma} = 0, \qquad
 \delta(\gamma+\hat \gamma - \lambda) =0.\label{b4.13}
\end{gather}

In order to obtain all symmetry algebras with $\dim L=2$ we must f\/ind all solutions of the system~(\ref{b4.13}). From the last equation we obtain either $\delta=0$, or $\lambda= \gamma+ \hat \gamma$. Thus the problem immediately branches in two. We then obtain $\hat \gamma$, $\hat \delta$ and $\omega$ from the f\/irst 3 equations. The remaining two equations provide nonlinear constraints on the remaining parameters. We shall not present the rather boring (computer assisted) analysis here, and only list the results. In each case we must make sure that we also have $r_2=3$ for the rank of $\hat M_2$.
\begin{alignat}{3}
& 1. \quad &&
K  =  \left(
\begin{array}{ccc}
1 & 0 & \gamma \\
\hat \beta & \lambda & 0  \\
\gamma + \hat \beta^2   &  \hat \beta (\lambda+\gamma) & (\lambda+\gamma)^2
\end{array}\right), \qquad (\hat \beta,\gamma, \lambda) \ne (0,0,0), & \nonumber\\
&&& \dot u_n =  \{(u_n^2+\gamma)u_{n+1}u_{n-1}+(\hat \beta u_n^2 + \lambda u_n)(u_{n+1}+u_{n-1}) + \gamma  u_n^2&\nonumber  \\
 &&& \phantom{\dot u_n =}{}  +  ( \hat \beta u_n + \lambda+\gamma)^2\}/(u_{n+1}-u_{n-1}), & \nonumber\\
&&& X  =  [\lambda+\gamma +\hat \beta u_n + u_n^2] \partial_{u_n}.& \nonumber
\\
& 2. \quad && K  =  \left(
\begin{array}{ccc}
1 & 0 & \gamma \\
0 & 2 \gamma & \delta  \\
\gamma   &  \delta & -3\gamma^2
\end{array}\right), \qquad (\gamma, \delta) \ne (0,0), \qquad \delta^2+4\gamma^3=0,\qquad  \gamma<0, & \nonumber \\
&&&  \dot u_n =  \frac{(u_n^2+\gamma)u_{n+1}u_{n-1}+(2 \gamma u_n+\delta)(u_{n+1}+u_{n-1}) + \gamma  u_n^2 + 2 \delta  u_n -3 \gamma^2}{u_{n+1}-u_{n-1}}, & \nonumber\\
&&& X  =  \frac{\gamma}{2 \delta}\left[\gamma - \frac{1}{2} u_n + u_n^2\right] \partial_{u_n} + t \partial_t. & \label{b4.16}
\\
& 3. \quad &&
K  =  \left(
\begin{array}{ccc}
1 & 0 & \gamma \\
\hat \beta & \lambda & \delta  \\
\hat \gamma   &  \hat \delta & \omega
\end{array}\right), \qquad (\hat \beta, \delta) \ne (0,0), \qquad \gamma<0,&  \nonumber \\
&&&  \hat \beta  =  2 \epsilon \sqrt{-\gamma} - \frac{\delta}{\gamma}, \qquad \epsilon=\pm 1, \qquad
\hat \gamma  =  - 3 \gamma + 2 \epsilon \frac{\delta}{\sqrt{-\gamma}}, & \nonumber\\
&&& \hat \delta  =  2 ( \delta + \epsilon (-\gamma)^{3/2}), \qquad
\omega  =  \gamma^2 - 6 \epsilon \sqrt{-\gamma}\delta, & \nonumber\\
&&& \dot u_n  =  \frac{(u_n^2+\gamma)u_{n+1}u_{n-1}+(\hat \beta u_n^2 + \lambda u_n+\delta)(u_{n+1}+u_{n-1}) + \hat \gamma  u_n^2 + \hat \delta  u_n + \omega}{u_{n+1}-u_{n-1}}, & \nonumber \\
&&& X  =  t \partial_t + \left[(\lambda+\gamma)\frac{\gamma}{2 \delta}+ \frac{1}{2}\left(\hat \beta \frac{\gamma}{\delta}-1\right)  u_n +\frac{ \gamma}{2 \delta} u_n^2\right] \partial_{u_n}. & \nonumber
\\
& 4.\quad  &&
K  =  \left(
\begin{array}{ccc}
1 & 0 & 0 \\
\hat \beta &  \lambda & 0  \\
\hat \beta^2  &  \hat \beta \lambda  & \lambda^2
\end{array}\right), \qquad \hat \beta \ne 0, \qquad \lambda \ne \hat \beta^2, & \nonumber \\
&&& \dot u_n  =  \frac{u_n^2 u_{n+1}u_{n-1}+ u_n (\hat  \beta u_n+\lambda)(u_{n+1}+u_{n-1}) + \hat \beta^2  u_n^2 +2 \hat \beta \lambda   u_n + \lambda^2}{u_{n+1}-u_{n-1}}, & \nonumber \\
&&& X  =  \big[\hat \beta \lambda +\hat \beta u_n + u_n^2\big] \partial_{u_n}.\nonumber
\\
& 5. \quad &&
K  =  \left(
\begin{array}{ccc}
1 & 0 & 0 \\
\hat \beta & 0 & 0 \\
0  &  0 & 0
\end{array}\right), \qquad \hat \beta \ne 0, & \nonumber \\
&&& \dot u_n  =  \frac{u_n^2u_{n+1}u_{n-1}+\hat \beta u_n^2 (u_{n+1}+u_{n-1}) }{u_{n+1}-u_{n-1}},\qquad
  X  =  -\left[ u_n + \frac{1}{2 \hat \beta} u_n^2\right] \partial_{u_n} + t \partial_t.\nonumber
\\
& 6. \quad &&
K  =  \left(
\begin{array}{ccc}
1 & 0 & 0\\
\hat \beta & \hat \beta^2 & 0  \\
\hat \beta^2  &  \hat \beta^3 & \hat \beta^4
\end{array}\right), \qquad \hat \beta \ne 0,& \nonumber \\
&&& \dot u_n  =  \frac{u_n^2u_{n+1}u_{n-1}+ \hat \beta ( u_n^2 +\hat \beta u_n)(u_{n+1}+u_{n-1}) +\hat \beta^2(  u_n^2 + 2 \hat \beta  u_n + \hat \beta^2)}{u_{n+1}-u_{n-1}}, & \nonumber \\
&&& X  =  \big[\hat \beta^2 + \hat \beta u_n + u_n^2\big] \partial_{u_n}.\nonumber
\end{alignat}

\noindent
{\bf Branch 2.} $\alpha=0$, $\beta=1$, $\gamma=0$, $\beta + \hat \beta =0$, $\delta+\hat \delta=0$, $\hat \gamma + 2 \lambda=0$, $\omega=0$.
\begin{alignat*}{3}
& 7. \quad &&
K  =  \left(
\begin{array}{ccc}
0 & 1 & 0\\
-1 & \lambda & \delta  \\
- 2 \lambda  &  -\delta & 0
\end{array}\right), \qquad (\lambda, \delta) \ne (0,0), & \nonumber \\
&&& \dot u_n  =  \frac{u_n u_{n+1}u_{n-1}+  ( -u_n^2 +\lambda u_n + \delta)(u_{n+1}+u_{n-1}) -2 \lambda  u_n^2 - 2 \delta  u_n }{u_{n+1}-u_{n-1}}, & \nonumber\\
&&& X  =  \big[\delta + \lambda u_n + u_n^2\big] \partial_{u_n}. & 
\end{alignat*}

\noindent
{\bf Branch 3.} $\alpha=0$, $\beta=0$, $\gamma=1$, $\hat \beta=\delta=\hat \delta=\omega=0$, $\hat \gamma = 1$, $\lambda=-1$.

We have $r_1=1$, $r_2=3$, so $\dim L=3$.

\subsubsection*{II. $\boldsymbol{r_1=4}$, $\boldsymbol{r_2=2}$}

We again follow the 3 branches.

\noindent
{\bf Branch 1.} $\alpha=1$, $\beta=0$.

{\bf I1.} $\gamma \ne 0$.
 We have
\begin{gather*} 
\hat M_2  \sim  \left(
\begin{array}{ccc}
1 & 0 & \lambda-\gamma \\
0 & 1 & -\frac{\delta}{\gamma}
\end{array}\right)
\end{gather*}
and all other entries in the row reduced matrix $\hat M_2$ must vanish (because $r_2=2$). We obtain
\begin{gather}
 \hat \gamma - \gamma + \hat \beta \frac{\delta}{\gamma} = 0, \qquad - \hat \delta + \hat \beta (\lambda -  \gamma)+ \hat \gamma \frac{\delta}{\gamma} = 0,
 \nonumber \\
   \omega - (\lambda - \gamma)^2 - \frac{\delta^2}{\gamma} = 0, \qquad \omega - (\lambda - \gamma)(\lambda - \hat \gamma) - \hat \delta \frac{\delta}{\gamma} = 0, \nonumber   \\
  (\gamma - \lambda)(\delta - \hat \delta) = 0, \qquad \hat \beta = 0, \qquad \delta - \hat \delta + \hat \beta(\lambda - \gamma) = 0. \label{b4.23}
\end{gather}
Conditions (\ref{b4.23}) imply
\begin{gather*} 
\hat \beta=0,\qquad \delta=\hat \delta \ne 0, \qquad \gamma=\hat \gamma, \qquad \omega=(\lambda-\gamma)^2+\frac{\delta^2}{\gamma}.
\end{gather*}
The result is:
\begin{alignat}{3}
& 8. \quad &&
K  =  \left(
\begin{array}{ccc}
1 & 0 & \gamma\\
0 & \lambda & \delta  \\
\gamma   &  \delta & (\lambda-\gamma)^2+\frac{\delta^2}{\gamma^2}
\end{array}\right), &\nonumber\\
 &&& (2\lambda\gamma^2 +\delta^2, \delta[2\gamma-\lambda],\delta[2\gamma^3+\lambda^2\gamma-\lambda \gamma^2 +\delta^2]) \ne (0,0,0),& \nonumber\\
&&& \dot u_n  =  \frac{(u_n^2+\gamma) u_{n+1}u_{n-1}+  (\lambda u_n + \delta)(u_{n+1}+u_{n-1}) +\gamma  u_n^2 + 2 \delta  u_n+(\lambda-\gamma)^2+\frac{\delta^2}{\gamma^2} }{u_{n+1}-u_{n-1}}, & \nonumber \\
&&& X  =  (-1)^n\left[\gamma-\lambda + \frac{\delta}{\gamma} u_n + u_n^2\right] \partial_{u_n}.\label{b4.25}
\end{alignat}

{\bf I2.} $\gamma = 0$.
To have $r_2=2$ we must put $\hat \beta = \delta=0$ and also $\hat \gamma=\hat \delta=0$. Then we obtain
\begin{alignat}{3}
& 9. \quad &&
K  =  \left(
\begin{array}{ccc}
1 & 0 & 0\\
0 & \lambda & 0  \\
0   &  0 & \omega
\end{array}\right), \qquad \omega \ne \lambda^2, &  \nonumber \\
&&& \dot u_n  =  \frac{u_n^2 u_{n+1}u_{n-1}+  \lambda u_n (u_{n+1}+u_{n-1}) +\omega }{u_{n+1}-u_{n-1}}, \qquad X  =  (-1)^n u_n \partial_{u_n}. &\label{b4.26}
\end{alignat}

The branches {\bf II} ($ \beta=1$, $\alpha=\gamma= \beta + \hat \beta = \delta + \hat \delta =  \hat \gamma + 2 \lambda = \omega =0$) and {\bf III} ($\alpha=\beta= \hat \beta = \delta =\hat \delta = \gamma + \lambda=\hat \gamma +  \lambda = \omega =0$, $\gamma=1$) do not yield any new result.

\section{Conclusions}\label{section5}

What we mean by ``integrable'' was def\/ined in the introduction. Thus an equation of the EYdKN family is integrable if and only if it satisf\/ies (\ref{e2}), i.e.\ it is of the YdKN type.

It follows from the previous analysis that the symmetry algebra $L$ of the EYdKN equation satisf\/ies $1 \le \dim L \le 5$. The largest dimension, namely~5, is achieved for the equation~(\ref{b4.3}). This is an integrable equation and in addition to point symmetries it allows higher symmetries. The two equations with four-dimensional Lie algebras, (\ref{b4.4}) and (\ref{b4.11}), are also both integrable. Of the three equations with three-dimensional symmetry algebras, (\ref{b4.5}) and (\ref{b4.5a}) are integrable but~(\ref{b4.8}) is not. Among the nine equations with two dimensional symmetry algebras only~(\ref{b4.16}),~(\ref{b4.25}) and~(\ref{b4.26}) are integrable (they possess higher symmetries) for all values of the parameters involved.

We see that integrable equations, i.e.\ those in the YdKN class, rather than in the EYdKN one, tend to have larger Lie point symmetry algebras than the nonintegrable ones. This is however not a reliable integrability criterion. Indeed the nonintegrable equation~(\ref{b4.8}) has a~three-dimensional symmetry algebra whereas the generic integrable equation in the class YdKN class with $\alpha \ne 0$ has only the one symmetry $X_0=\partial_t$ (specif\/ically an equation with $\alpha=1$, $\beta=0$, $\lambda \ne 2 \gamma$, $\omega \ne  \gamma ( \lambda -\gamma^2 + \delta^2)$).

This rather loose relation between Lie point symmetries and integrability was already observed in a symmetry analysis of Toda type equations~\cite{r3}.

A complete symmetry analysis of the integrable GYdKN equation (\ref{e2*}), (\ref{e2**}) is not attempted here. We will just present one non-trivial example. Work is in progress to provide a complete classif\/ication.

 An example of the GYdKN equation is
\begin{gather*} 
\dot u_n  =  \frac{\chi_{n+1}(u_{n+1}+u_{n-1})+2 \chi_n u_n}{u_{n+1}-u_{n-1}}, \qquad
\chi_n  = \frac{1+(-1)^n}{2}, \qquad \chi_{n+1} =\frac{1-(-1)^n}{2}.
\end{gather*}
Using the same approach as in Sections~\ref{section3} and~\ref{section4} above, we f\/ind that the symmetry algebra is four-dimensional with basis
\begin{gather*} 
X_0  = \partial_t, \qquad X_1 =t \partial_t + u_n \partial_{u_n}, \qquad
X_2  = \chi_{n+1} \partial_{u_n}, \qquad  X_3 = \chi_n u_n \partial_{u_n}.
\end{gather*}
This is a direct sum
$
\{ X_0, X_1, X_2\}+\{X_3\}
$,
where $\{ X_0, X_1, X_2\}$ is solvable with $\{ X_0,  X_2\}$ as its Abelian nilradical.

The limit from the discrete equations considered in this article to the usual (continuous) Krichever--Novikov equation is quite complicated (see Section~\ref{section2})  and  does not preserve symmetries. From Table~\ref{table1} we see that the largest symmetry algebra is obtained for $f(u)=0$ and satisf\/ies $\dim L =6$. Equation~(\ref{c1}) in this case remains nontrivial.  It is just  the Schwarzian KdV equation \cite{glz_2004,nhn2000,weiss83}. A discrete analogue in this case would be $P=Q=R = 0$ in~(\ref{4.1}),~(\ref{4.2}). This equation is trivial, the symmetry algebra is inf\/inite-dimensional generated by
\begin{gather*} 
X(\tau)=\tau(t) \partial_t, \qquad U(\phi_n)= \phi_n (u_n)\partial_{u_n},
\end{gather*}
where $\tau(t)$ and $\phi_n(u_n)$ are arbitrary ($C^{\infty}$) functions of their arguments.

The Lie algebra element $P_1=\partial_x$, generating space translations, is always absent in the discrete case. Formally we can add the operator $\hat N = \partial_n$ to the symmetry algebra, as was done previously for the Toda lattice~\cite{r4,r3}. This corresponds (formally) to introducing a (discrete) group transformation $n^* = n + N$ with the understanding that $N$ is an integer (a shift on the lattice). This symmetry allows us to consider a periodic EYdKN equation, or equivalently to restrict to a f\/inite lattice.

Finally, let us just give some examples showing how the Lie point symmetries can be used to reduce the considered dif\/ferential-dif\/ference equation to simpler equations. Consider  (\ref{b4.3}) and its dilation subalgebra $X_1$. A solution invariant under the subgroup generated by $X_1$ will have the form $u_n= c_n t^{-1/2}$. Putting this into (\ref{b4.3}) we f\/ind that the coef\/f\/icient $c_n$ must satisfy the nonlinear dif\/ference equation
\begin{gather*} 
 c_n =  \frac{1}{2} \left[\frac{1}{c_{n+1}} - \frac{1}{c_{n-1}} \right], \qquad u_n = \frac{c_n}{\sqrt{t}}.
\end{gather*}
Similarly the subalgebra $ X_3+a X_0$ leads to the invariant solution
\begin{gather*} 
u_n = \frac{a}{c_n-t}, \qquad c_n=- \frac{a^2}{2} n + \beta.
\end{gather*}

 Work is in progress for  a complete study of the group invariant solutions for all the invariant equations obtained in this article.

\subsection*{Acknowledgements}

The research of L.D.\  has been partly supported by the Italian Ministry of Education and Research,  PRIN ``Continuous and discrete nonlinear integrable evolutions: from water
waves to symplectic maps''.
The research of P.W.\ was partly supported by a research grant from NSERC of Canada.
R.I.Y.\ has been partially supported by the Russian Foundation for Basic Research (grant
numbers 10-01-00088-a and 11-01-97005-r-povolzhie-a).

\pdfbookmark[1]{References}{ref}
\LastPageEnding

\end{document}